\documentclass[fleqn]{elsart}
\usepackage{amssymb}
\usepackage{indentfirst}
\usepackage{psfig,color}
\usepackage{epsfig}
\usepackage{epsf}
\usepackage{graphicx}

\newcommand{\kp}{\mathbf{k}_\perp}
\newcommand{\p}{\perp}
\newcommand{\ssh}{\!\!\!/}
\journal{Physics Letters B}

\begin{document}

\begin{frontmatter}
\title{Azimuthal asymmetry in unpolarized $\pi N$ Drell-Yan process}

\author[pku]{Zhun Lu},
\author[ccast,pku,it]{Bo-Qiang Ma\corauthref{cor}}
\corauth[cor]{Corresponding author.} \ead{mabq@phy.pku.edu.cn}
\address[pku]{Department of Physics, Peking University, Beijing 100871, China}
\address[ccast]{CCAST (World Laboratory), P.O.~Box 8730, Beijing
100080, China}
\address[it]{Di.S.T.A., Universit\`a del Piemonte Orientale
``A. Avogadro'' and INFN, Gruppo Collegato di Alessandria, 15100
Alessandria, Italy}

\begin{abstract}
Taking into account the effect of final state interaction, we
calculate the non-zero (na\"{i}ve) $T$-odd transverse momentum
dependent distribution $h_1^{\perp}(x,\kp^2)$ of the pion in a
quark-spectator-antiquark model with effective
pion-quark-antiquark coupling as a dipole form factor. Using the
model result we estimate the $\cos 2\phi$ asymmetries in the
unpolarized $\pi^- N$ Drell-Yan process which can be expressed as
$h_1^{\perp}\times\bar{h}_1^{\perp}$. We find that the resulting
$h_{1\pi}^\p(x,\kp^2)$ has the advantage to reproduce the
asymmetry that agrees with the experimental data measured by NA10
Collaboration. We estimate the $\cos2\phi$ asymmetries averaged
over the kinematics of NA10 experiments for 140, 194 and 286~GeV
$\pi^-$ beam and compare them with relevant experimental data.
\end{abstract}
\end{frontmatter}

\section{Introduction}

Recently it is demonstrated that the effect of final state
interaction (FSI) or initial state interaction (ISI) can lead to
significant azimuthal asymmetries in various high energy
scattering processes involving hadrons~\cite{bhs02a,bhs02b}. Among
these asymmetries, single spin asymmetry (SSA) in semi-inclusive
deeply inelastic scattering (SIDIS)~\cite{bhs02a} and that in
Drell-Yan processes~\cite{bhs02b} from FSI/ISI via the exchange of
a gluon, have been explored and are recognized as previously known
Sivers effect~\cite{sivers90,abm95}. This effect, formerly thought
to be forbidden by the time-reversal property of
QCD~\cite{collins93}, can be survived from time-reversal
invariance due to the presence of the path-ordered exponential
(Wilson line) in the gauge-invariant definition of the transverse
momentum dependent parton
distributions~\cite{collins02,bjy03,bmp03}. Along this direction
some phenomenological studies~\cite{gg03,bsy04,lm04a} have been
carried out on transverse single-spin asymmetries in SIDIS
process, which is under investigation by current
experiment~\cite{hermes04}. Analogously the exchange of a gluon
can also lead to another leading twist (naive) $T$-odd
distribution $h_1^\p(x,\kp^2)$: the covariant transversely
polarization density of quarks inside an unpolarized hadron. This
chiral-odd partner of Sivers effect function, introduced first in
Ref.~\cite{bm98} and is referred to as Boer-Mulders function, has
been proposed~\cite{boer99} to account for the large $\cos2\phi$
asymmetries in the unpolarized pion-nucleon Drell-Yan process that
were measured more than 10 years ago~\cite{na10,conway89}.
Recently $h_1^{\perp}(x,\kp^2)$ of the proton has been computed in
a quark-scalar diquark model~\cite{gg03,bbh03} and also used to
analyze the consequent $\cos 2\phi$ azimuthal asymmetries in both
unpolarized $ep$ SIDIS process~\cite{gg03} and unpolarized
$p\bar{p}$ Drell-Yan process~\cite{bbh03} respectively.

The same mechanism producing $T$-odd distribution functions can be
applied to other hadrons such as mesons. In a previous
paper~\cite{lm04b} we reported that non-zero $h_1^\p$ of the quark
inside the pion (denoted as $h_{1\pi}^\p$) can also arise from
final state interaction, by applying a simple quark
spectator-antiquark model. Among the phenomenological implications
of the function $h_{1\pi}^\p$ is an important result for the
$\cos2\phi$ azimuthal asymmetry in the unpolarized $\pi^-N$
Drell-Yan process~\cite{na10,conway89}, which can be produced by
the product of $h_1^{\perp}$ of the pion and that of the nucleon.
Therefore, one can investigate how the theoretical prediction of
the asymmetry is comparable with the experimental result, as a
test of the theory and the model. In the present paper, based on
$h_{1\pi}^\p$ from our model calculation, we analyze the
$\cos2\phi$ azimuthal asymmetry in the unpolarized $\pi^- N$
Drell-Yan process by considering the kinematical region of NA10
experiments~\cite{na10}. To obtain the right $Q_T$ dependence of
the asymmetry, we recalculate $h_{1\pi}^\p(x,\kp^2)$ in a
spectator model similar to the model used in Ref.~\cite{lm04b}.
The difference is that here we treat the effective
pion-quark-antiquark coupling $g_\pi$ as a dipole form factor
$g_\pi(k^2)$, in contrary to the treatment in Ref.~\cite{lm04b}
where we take $g_\pi$ as a constant. We find that
$h_{1\pi}^\p(x,\kp^2)$ resulting from the new treatment together
with $h_1^{\perp}(x,\kp^2)$ for the nucleon in a similar
treatment~\cite{bsy04} can reproduce the $\cos2\phi$ asymmetry
which agrees with NA10 data. We give the asymmetries predicted by
our model averaged over the kinematics of NA10 experiments for
140, 194 and 286~GeV $\pi^-$ beam and find that the energy
dependence of these asymmetries is not strong.

\section{Non-zero $h_{1\pi}^\p$ of the pion in
spectator model}

In this section, we will show how to calculate
$h_{1\pi}^\p(x,\kp^2)$ in a quark-spectator antiquark model. We
follow Ref.~\cite{bhs02a} to work in Abelian case at first and
then generalize the result to QCD. There are pion-quark-antiquark
interaction and gluon-spectator antiquark interaction in the
model:
\begin{equation}
\mathcal{L}_I=-g_\pi\bar{\psi}\gamma_5\psi\varphi_\pi-e_2\bar{\psi}\gamma^\mu\psi
A_\mu+h.c.,
\end{equation}
in which $g_\pi$ is the pion-quark-antiquark effective coupling,
and $e_2$ is the charge of the antiquark. When the intrinsic
transverse momentum of the quark is taken into account, as
required by $T$-odd distributions, the quark correlation function
of the pion in Feynman gauge (we perform calculation in this
gauge) is~\cite{bjy03,bmp03}:
\begin{eqnarray}
\Phi_{\alpha\beta}(x,\kp)=\int \frac{d \xi^- d^2
\mathbf{\xi}_\perp}{(2\pi)^3}e^{ik\cdot\xi}\langle
P_\pi|\bar{\psi}_\beta(0) \mathcal{L}_0(0^-,\infty^-)
\mathcal{L}_\xi^\dag(\xi^-,\infty^-)\psi_\alpha(\xi)|P_\pi\rangle\bigg{|}_{\xi^+=0},\label{phi}
\end{eqnarray}
where $\mathcal{L}_a(a^-,\infty^-)$ is the path-ordered
exponential (Wilson line) accompanied with the quark field which
has the form
\begin{equation}
\mathcal{L}_0(0,\infty)=\mathcal{P}~\textmd{exp}\left
(-ig\int_{0^-}^{\infty^-}
A^+(0,\xi^-,\mathbf{0}_\perp)d\xi^-\right),
\end{equation}
etc. The Wilson line has the importance to make the definition of
the distribution/correlation function gauge-invariant. Without the
constraint of time-reversal invariance, in leading twist the quark
correlation function of the pion can be parameterized into a set
of leading twist transverse momentum dependent distribution
functions as follows~\cite{bm98,tm96}
\begin{equation}
\Phi(x,\kp)=\frac{1}{2}\left
[f_{1\pi}(x,\kp^2)n\ssh+h_{1\pi}^\perp(x,\kp^2)\frac{\sigma_{\mu\nu}\kp^\mu
n^{\nu}}{M_\pi}\right ],
\end{equation}
where $n$ is the light-like vector with components $(n^+, n^-,
\mathbf{n}_\perp)=(1, 0, \mathbf{0}_\perp)$,
$\sigma_{\mu\nu}=\frac{i}{2}[\gamma_\mu,\gamma_\nu]$ and $M_\pi$
is the pion mass. Knowing $\Phi_\pi(x,\kp)$, one can obtain these
distributions from equations
\begin{eqnarray}
&f_{1\pi}(x,\kp^2)&=\textmd{Tr}[\Phi(x,\kp)\gamma^+];\\
&\frac{2h_{1\pi}^{\perp}(x,\kp^2)\mathbf{k}_\p^i}{M_\pi}&=
\textmd{Tr}[\Phi(x,\kp)\sigma^{i+}].\label{phitoh1t}
\end{eqnarray}

\begin{figure}
\begin{center}
\scalebox{0.8}{\includegraphics{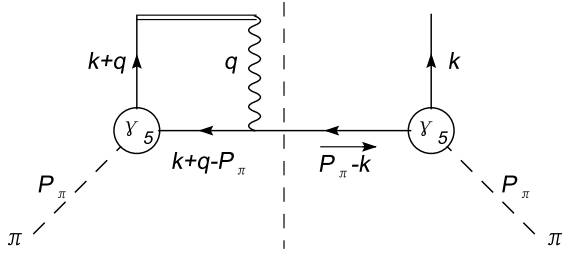}}\caption{\small
Effective correlation function $\Phi$ in the antiquark spectator
model with final-state interaction modelled by one gluon
exchange.}\label{fsi}
\end{center}
\end{figure}

The calculation of unpolarized distribution function $f_{1\pi}$ in
the antiquark spectator model can be done~\cite{jmr97} from the
lowest order (without Wilson line) correlation function in
Eq.~(\ref{phi}). However it can not lead to any $T$-odd
distribution function such as $h_{1\pi}^\p$. As demonstrated in
Ref.~\cite{bhs02a}, the non-zero $T$-odd distribution requires
final state interaction from gluon exchange between the struck
quark and target spectator. Here we follow the observation in
Refs.~\cite{collins02} and \cite{jy02} that final state
interaction in an initial hadron state can be taken into account
effectively by introducing an appropriate Wilson line in the
gauge-invariant definition of the transverse momentum dependent
distribution function, or equivalently, quark correlation function
of the hadron. The Wilson line can provide nontrivial phase needed
for T-odd distribution functions. Since we have defined such a
correlation function in Eq.~(\ref{phi}), we can start from
Eq.~(\ref{phi}) to calculate $h_{1\pi}^\p$ with the explicit
presence of the Wilson line. We expand the Wilson line to first
order corresponding to one gluon exchange. Therefore, according to
Eq.~(\ref{phitoh1t}) and Eq.~(\ref{phi}), $h_{1\pi}^\p$ can be
calculated in the antiquark spectator model from the expression
\begin{eqnarray}
&&\frac{2h_{1\pi}^\p(x,\kp^2)\kp^i}{M_\pi}=\sum_{\bar{q}^{s}}\frac{1}{2}
\int\frac{d\xi^-d\xi_\p}{(2\pi)^3} e^{ik\cdot\xi}\langle
P_\pi|\bar{\psi}_\beta(0)|\bar{q}^{s}\rangle\nonumber\\
&&\times\langle \bar{q}^{s}|\left(-ie_1 \int_{\xi^-}^{\infty^-}
A^+(0,\xi^-,\mathbf{0}_\p)d\xi^-\right)
\sigma^{i+}_{\beta\alpha}\psi_{\alpha}(\xi)|P_\pi\rangle\bigg{|}_{\xi^+=0}+h.c.,
~~~~~~ \label{exp}
\end{eqnarray}
in which $|\bar{q}^s\rangle$ represents the antiquark spectator
state with spin $s$, and $e_1$ is the charge of the struck quark.

Fig.~\ref{fsi} is the diagram equivalent to Eq.~(\ref{exp}). The
figure shows the effective correlation function in the antiquark
spectator model with the Wilson line expanding to the first order.
$h_{1\pi}^\p$ can be obtained from the diagram by inserting
$\sigma^{i+}$, according to Eq.~(\ref{phitoh1t}). Fig.~\ref{fsi}
is similar to the diagram used by Ji and Yuan~\cite{jy02} to
calculate $f_{1T}^\p$ of the proton in scalar diquark model. The
$\gamma_5$ inside the circle denotes that the pion-quark-antiquark
coupling is pseudoscalar coupling. The double line represents the
eikonalized quark propagator (eikonal line), which is produced by
the Wilson line along the light-cone vector
$\bar{n}^\mu=(\bar{n}^+,\bar{n}^-,\bar{\mathbf{n}}_\perp)=(0,1,\mathbf{0}_\perp)$.
The eikonal line gives rise the final state interaction effect
between the fast moving struck quark and the gluon field from
target spectator system~\cite{collins02,jy02}. The Feynman rule
for the eikonal line is $1/(q^++i\epsilon)$~\cite{cs82} (see also
Appendix in Ref.~\cite{bbh03}), where $q$ is the momentum of the
gluon attached to the eikonal line. The Feynman rule for the
eikonal line-gluon vertex is $e_1\bar{n}^\mu$~\cite{cs82}. The
straight line cut by the vertical dashed line denotes the on-shell
spectator antiquark state $v^{s}$ or $\bar{v}^{s}$.

Usually there are two choices of the pion-quark-antiquark coupling
$g_\pi$:
\begin{itemize}
    \item case 1 : $g_\pi$ as a normalization constant which is used in Ref.~\cite{lm04b}.
    A similar treatment for the proton-quark-diquark coupling $g$ has been
    adopted in Refs.~\cite{bhs02a,bbh03} to estimate $f_{1T}^\p(x,\kp^2)$
     and $h_{1}^\p(x,\kp^2)$ of the proton.
    \item case 2 : $g_\pi$ as a dipole form factor~\cite{jmr97}
\begin{equation}
g_\pi(k^2)=N_\pi\frac{k^2-m^2}{(\Lambda^2-k^2)^2}=N_\pi(1-x)^2\frac{k^2-m^2}
{(\kp^2+L_\pi^2)^2},
\end{equation}
with
\begin{eqnarray}
L_\pi^2&=&(1-x)\Lambda^2+xm^2-x(1-x)M_\pi^2,\\
\kp^2&=&-(1-x)k^2-xm^2+x(1-x)M_\pi^2,
\end{eqnarray}
and $N_\pi$ is the normalization constant, $m$ is the mass of the
quark/antiquark inside the pion, $\Lambda$ is the cut off
parameter of the quark momentum. This kind of treatment has been
applied to model T-even nucleon distribution
functions~\cite{jmr97}, and recently in the calculations of
$f_{1T}^\p(x,\kp^2)$ and $h_{1}^\p(x,\kp^2)$ of the
proton~\cite{bsy04} in order to eliminate the divergences in the
$k_\p$-moments of these $k_\p$-dependent distribution functions.
\end{itemize}

In the previous paper~\cite{lm04b} we performed a computation on
$h_{1\pi}^\p$ in case 1 which yields:
\begin{equation}
h_{1\pi}^\p(x,\kp^2)=\frac{A_{\pi}}{\kp^2(\kp^2+B_\pi)}\textmd{ln}\left
(\frac{\kp^2+B_\pi}{B_\pi}\right ) \label{h1p},
\end{equation}
and the corresponding unpolarized distribution is
\begin{equation}
f_{1\pi}(x,\kp^2)=C_{\pi}\frac{\kp^2+D_{\pi}}{(\kp^2+B_{\pi})^2},
\end{equation}
where
\begin{eqnarray}
&&A_\pi=\frac{g_\pi^2}{2(2\pi)^3}
\frac{|e_1e_2|}{4\pi}mM_\pi(1-x),
~~~~~B_\pi=m^2-x(1-x)M_\pi^2, \\
&&C_\pi=(1-x)g_\pi^2/[2(2\pi)^3], ~~~~~~~~~~~~~~D_\pi=(1+x)^2m^2.
\end{eqnarray}
An interesting result is that the transverse momentum dependence
of $h_{1\pi}^\p(x,\kp^2)$ in this model is the same as that of
$h_1^{\p}$ of the proton in the quark-scalar diquark
model~\cite{bbh03}.

Now we perform the computation of $h_{1\pi}^\p$ in case 2, that
is, in the situation of $g_\pi$ as a dipole form factor. According
to Eq.~(\ref{exp}), also with the help of Fig.~\ref{fsi} and the
Feynman rules introduced above, we can calculate $h_{1\pi}^\perp$
from the integral:
\begin{eqnarray}
&&\frac{2h_{1\pi}^\p(x,\kp^2)\kp^i}{M_\pi}=\frac{i|e_1e_2|}{8(2\pi)^3(1-x)P_\pi^+}
\sum_{s}\int\frac{d^4q}{(2\pi)^4}\bar{v}^sg_\pi(k^2)\gamma_5\frac{k\ssh+m}{k^2-m^2}
\sigma^{i+}\nonumber\\
&&\times\frac{k\ssh+q\ssh+m}{(k+q)^2-m^2}g_\pi((k+q)^2)\gamma_5
\frac{k\ssh+q\ssh-P\ssh\!_\pi+m}{(k+q-P_\pi)^2-m^2+i\epsilon}\nonumber\\
&&\times\gamma^+v^s\frac{1}{q^++i\epsilon}\frac{1}{q^2-i\epsilon}+h.c.
. \label{h1t}
\end{eqnarray}
In above equation we have used $\langle
P_\pi|\bar{\psi}(0)|\bar{q}^{s}\rangle=\bar{v}^{s}g_\pi(k^2)
\gamma_5 i(k\ssh+m)/(k^2-m^2)$, etc, which is a result of the
spectator model~\cite{jmr97}. The $\gamma^+$ in the last line of
Eq.~(\ref{h1t}) comes from the contraction of the eikonal
line-gluon vertex and the gluon-antiquark vertex: $\bar{n}^\mu
g_{\mu\nu}\gamma^\nu=\gamma^+$. The loop integral over the gluon
momentum $q$ is similar to the integral for calculating
$f_{1T}^\perp$ in \cite{bbh03} and \cite{jy02}. The $q^-$ integral
is realized from contour method, and $q^+$ integral can be done by
taking the imaginal part of the eikonal propagator:
$1/(q^++i\epsilon)\rightarrow -i\pi\delta(q^+)$, since the real
part of the propagator is cancelled by the Hermitian conjugate
term. After performing the integral we yield $h_{1\pi}^\p$ with a
form different from Eq.~(\ref{h1p}):
\begin{equation}
h_{1\pi}^\p(x,\mathbf{k}_\p^2)=\frac{|e_1e_2|}{4\pi}
\frac{N_\pi^2(1-x)^3mM_\pi}{2(2\pi)^3L_\pi^2(\mathbf{k}_\p^2+L_\pi^2)^3}.\label{h1pi}
\end{equation}
The corresponding unpolarized distribution is
\begin{equation}
f_{1\pi}(x,\kp^2)=\frac{N_\pi^2(1-x)^3(\kp^2+D_{\pi})}{2(2\pi)^3(\mathbf{k}_\p^2+L_\pi^2)^4}.
\end{equation}

To calculate the trace in the nominator of Eq.~(\ref{h1t}) we take
the spin sum of the antiquark state as $\sum_s
v^s\bar{v}^s=(P\ssh_\pi-k\ssh-m)$, which is a little different
from the spin sum adopted in Ref.~\cite{jmr97}. One can find that
the form of Eq.~(\ref{h1pi}) is similar to $h_1^\p$ of the proton
computed in Ref.~\cite{bsy04}. We also calculate
$\bar{h}_{1\pi}^\p$, the $T$-odd distribution of the valence
antiquark inside the pion, and yield
$\bar{h}_{1\pi}^\p=h_{1\pi}^\p$. Comparing the two versions of
$h_{1\pi}^\p$ in Eq.~(\ref{h1pi}) and Eq.~(\ref{h1p}), we find
that each one has a significant magnitude, which means both of
them can give unsuppressed $\cos2\phi$ asymmetry. However, the
transverse momentum dependence of the two versions are very
different, that is to say, the $Q_T$ behavior of the $\cos2\phi$
asymmetry predicted by the two cases should be different. One may
expect experiments to make a discrimination between the two
versions of $h_{1\pi}^\p$. We will give a further comparison with
available experimental data in next section.

\section{The $\cos 2\phi$ asymmetry in the unpolarized $\pi^-N$ Drell-Yan
process} 
\begin{figure}
\begin{center}
\scalebox{0.8}{\includegraphics{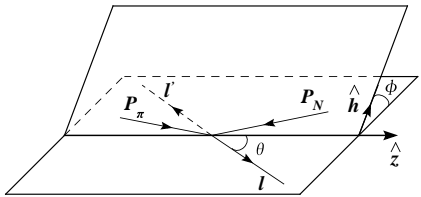}}\caption{\small  Angular
definitions of the unpolarized Drell-Yan process in the lepton
pair center of mass frame.}\label{drell-yan}
\end{center}
\end{figure}

The unpolarized Drell-Yan process cross section has been
measured in muon pair production by pion-nucleon collision:
$\pi^-N\rightarrow\mu^+\mu^-X$, with $N$ denoting a nucleon in
deuterium or tungsten and a $\pi^-$ beam with energy of 140, 194,
286~GeV~\cite{na10} and 252~GeV~\cite{conway89}. The general form
of the angular differential cross section for the unpolarized
Drell-Yan process is
\begin{equation}
\frac{1}{\sigma}\frac{d\sigma}{d\Omega}=\frac{3}{4\pi}\frac{1}{\lambda+3}
\left (1+\lambda\cos^2\theta+\mu
\sin2\theta\cos\phi+\frac{\nu}{2}\sin^2\theta\cos2\phi\right ),
\end{equation}
where $\phi$ is the angle between the lepton plane and the plane
of the incident hadrons in the lepton pair center of mass frame
(see Fig.~\ref{drell-yan}). The definition of the lepton plane
depends on the choice of axes $\hat{z}$ in the lepton pair center
of mass system. In our calculation we choose $\hat{z}$ parallel to
the bisector of $\vec{\mathbf{P}}_\pi$ and $-\vec{\mathbf{P}}_N$,
which is referred to as Collins-Soper frame~\cite{cs77}. The
experimental data show large value of $\nu$ near to 30\% in the
Collins-Soper frame. The asymmetry predicted by perturbative QCD
is expected to be small~\cite{na10,bnm93}. Several theoretical
approaches have been suggested to interpret the experimental data,
such as high-twist effect~\cite{bbkd94,ehvv94} and factorization
breaking mechanism~\cite{bnm93}. A natural explanation has been
proposed by Boer~\cite{boer99} that the product of two $T$-odd
chiral-odd $h_1^\perp$ can give $\cos2\phi$ asymmetry without
suppression by the momentum of the lepton pair. In that paper, a
parametrization of $h_1^\perp$ in a similar form of Collins
fragmentation function~\cite{collins93} has been given to fit the
experiment data. Recently it is found that non-zero $h_1^\perp$
can arise from final state interaction without violation of
time-reversal invariance, and has been used to estimate
$\cos2\phi$ asymmetry in the unpolarized $p\bar{p}\rightarrow
l\bar{l}X$ Drell-Yan process~\cite{bbh03}.

Encouraged by this proposal, we apply $h_{1\pi}^\p$ given by our
model calculation to estimate the consequent $\cos 2\phi$
asymmetry in the unpolarized $\pi^- N$ Drell-Yan process measured
by NA10 Collaboration. In case the vector boson that produces the
lepton pair is a virtual photon, the leading order unpolarized
Drell-Yan cross section expressed in the Collins-Soper frame
is~\cite{boer99}
\begin{eqnarray}
&\frac{d\sigma(h_1h_2\rightarrow l\bar{l}X)}{d\Omega
dx_1dx_2d^2\mathbf{q}_\perp}&=
\frac{\alpha^2_{em}}{3Q^2}\sum_{a}\Bigg{\{}
A(y)\mathcal{F}[f_1^a\bar{f}_1^{\bar{a}}] +B(y)\nonumber\\
&&\times\cos2\phi~\mathcal{F}\left [(2\hat{\mathbf{h}}\cdot
\mathbf{p}_\perp\hat{\mathbf{h}}\cdot \mathbf{k}_\perp
-\mathbf{p}_\perp\cdot
\kp)\frac{h_1^{\perp,a}\bar{h}_1^{\perp,\bar{a}}}{M_1M_2}\right
]\Bigg{\}},\label{cs}
\end{eqnarray}
where $Q^2=q^2$ is the invariance mass square of the lepton pair,
$\mathbf{q}_\perp$ is the transverse momentum of the pair, and the
vector $\hat{\mathbf{h}}=\mathbf{q}_\perp/Q_T$. We have used the
notation
\begin{eqnarray}
\mathcal{F}[f_1\bar{f}_1]&=&\int d^2\mathbf{p}_\perp
d^2\kp\delta^2(\mathbf{p}_\perp+\kp-\mathbf{q}_\perp)
f_1(x,\mathbf{p}_\perp^2)\bar{f}_1(\bar{x},\kp^2).
\end{eqnarray}
From Eq.~(\ref{cs}) one can give the expression for the asymmetry
coefficient $\nu$~\cite{boer99}:
\begin{eqnarray}
\nu&=&2\sum_{a}e_a^2\mathcal{F}\left [(2\hat{\mathbf{h}}\cdot
\mathbf{p}_\perp\hat{\mathbf{h}}\cdot \mathbf{k}_\perp
-\mathbf{p}_\perp\cdot
\kp)\frac{h_1^{\perp,a}\bar{h}_1^{\perp,\bar{a}}}{M_1M_2}\right
]\Bigg{/}\sum_{a}e_a^2\mathcal{F}[f_1^a\bar{f}_1^{\bar{a}}].\label{nu}
\end{eqnarray}

\begin{figure}
\begin{center}
\scalebox{0.6}{\includegraphics{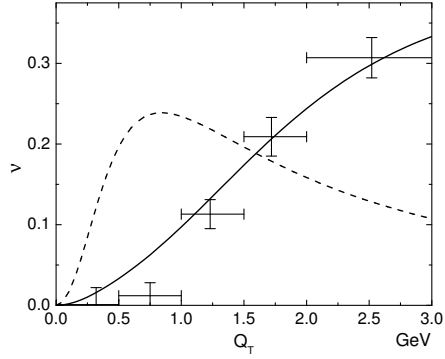}}\caption{\small The
$\cos2\phi$ asymmetry (solid line) in the unpolarized $\pi^-N$
Drell-Yan process defined in the Collins-Soper frame at
$\bar{x}=x=0.5$. The dashed line represents asymmetry given by the
previous model result~\cite{lm04b} (with $h_{1\pi}^\p$ calculated
in case 1). The data are taken from Ref.~\cite{na10} at 194~GeV.
}\label{cos2phi}
\end{center}
\end{figure}

The $\cos 2\phi$ dependence as observed by the NA10 Collaboration
does not show a strong dependence on $A$~\cite{na10}, {\it i.e.,}
the asymmetry is unlikely associated with nuclear effect. The
leading contribution which comes from the valence quarks is
$\bar{h}_{1\pi}^{\p,\bar{u}}\times h_1^{\p,{u}}$, therefore we can
adopt the $u$-quark dominance, {\it i.e.}, we do not include sea
quark contribution which is expected to be small. We use
$h_{1\pi}^\p$ given in Eq.~(\ref{h1pi}) (case 2) to estimate the
asymmetry. We also need $h_1^\p$ of the nucleon in a similar
treatment with effective coupling as a dipole form factor. This
has been done in Ref.~\cite{bsy04}. We use this version $h_1^\p$
but only include contribution from the scalar diquark
($h_1^{\p,u}=h_1^{\p, S}$, with $S$ denoting the scalar diquark).
There are two considerations: the first is to reduce the number of
the parameters, and the second is because of the $u$-quark
dominance assumption. Based on Eq.~(\ref{nu}) with the denominator
from the same model result, we give the numerical estimation of
the asymmetry at $\bar{x}=x=0.5$ in Fig.~\ref{cos2phi} (shown by
the solid curve) with experiment data from NA10 Collaboration. For
the parameters in the expressions of $h_{1\pi}^\p$ and $h_1^\p$ we
choose: $\Lambda=0.6~\textmd{GeV}$, $M_\pi=0.137~\textmd{GeV}$,
$m=0.1~\textmd{GeV}$, $M_N=0.94~\textmd{GeV}$,
$\lambda_S=0.8~\textmd{GeV}$, and $m_N=0.3~\textmd{GeV}$, where
$M_N$, $\lambda_S$ and $m_N$ are the nucleon mass, the scalar
diquark mass and the mass of the quark inside the nucleon,
respectively. For the coupling constant $|e_1e_2|/4\pi$ we
extrapolate $|e_1e_2|/4\pi\rightarrow C_F\alpha_s$, and take
$\alpha_s=0.3$ and $C_F=4/3$ which are adopted in
Ref.~\cite{bhs02a}. We choose the data at 194~GeV of
Ref.~\cite{na10}, since the error bars of them are smallest (the
error in $Q_T$ is chosen to be the bin size). We find that the
estimated asymmetry agrees with the experiment data fairly well,
although our estimation is crude since some approximations are
adopted. In contrast, as shown by the dashed line in
Fig.~\ref{cos2phi}, the $Q_T$ shape of the asymmetry produced by
$h_{1\pi}^\p$ denoted in Eq.~(\ref{h1p}) (case 1 with the
pion-quark-antiquark coupling as a constant) is not consistent
with experimental data (see Ref.~\cite{lm04b}).

\begin{figure}
\begin{center}
\scalebox{0.67}{\includegraphics{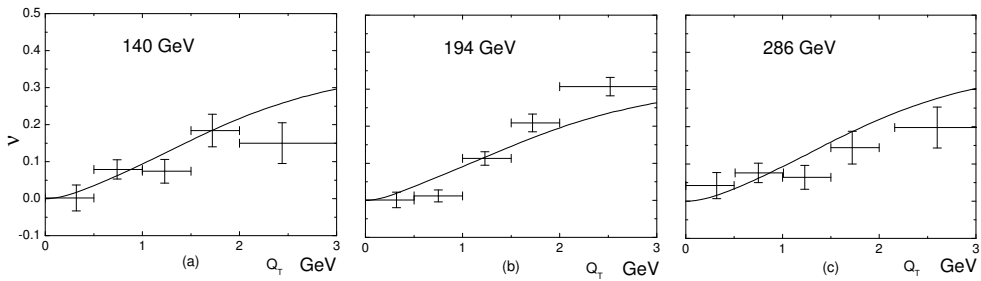}}\caption{\small The
$\cos2\phi$ asymmetries in the Collins-Soper frame for pion beam
with different energy averaged over kinematics region:
$\bar{x}<0.7$, $4.0\leq Q\leq 8.5~\textmd{GeV}$ ($4.05\leq Q\leq
8.5~\textmd{GeV}$ for the $194~\textmd{GeV}$ Data) and $Q \geq
11~\textmd{GeV}$. The data are taken from Ref.~\cite{na10}.
}\label{na10}
\end{center}
\end{figure}

We further use $h_{1\pi}^\p$ given in Eq.~(\ref{h1pi}) and the
same parameters adopted above to estimate the asymmetries averaged
over the kinematics of NA10 experiments. For $Q_T^2\ll Q^2$, the
momentum fractions of the quarks inside the pion and the nucleon
satisfy the relation: $\bar{x}x=Q^2/s$, where $\sqrt{s}$ is the
center of mass energy of the pion-nucleon system, for instance,
for 194~GeV beam $\sqrt{s}=\sqrt{(194+0.94)^2-194^2}=19.1$~GeV,
and for 140~GeV, 286~GeV beam $\sqrt{s}=16.2$~GeV, 23.2~GeV
respectively~\cite{na10}. The data-selecting condition of the NA10
experiments is: $\bar{x}<0.7$, $4.0\leq Q\leq 8.5~\textmd{GeV}$
($4.05\leq Q\leq 8.5~\textmd{GeV}$ for the $194~\textmd{GeV}$
Data) and $Q \geq 11~\textmd{GeV}$. We use above kinematical
constrains to evaluate the averaged asymmetries for $\pi^-$ beam
with different energy. In Fig.~\ref{na10} we plot the asymmetries
in the Collins-Soper frame versus $Q_T$ for 140~GeV, 194~GeV and
286~GeV beam together with the data of Ref.~\cite{na10}. The
estimated asymmetries for pion beam with different energy are
still consistent with experimental data. Our estimation shows that
the energy dependence of the asymmetries is not strong, and this
agrees with the experimental observation.

\section{Summary}
The observed large $\cos2\phi$ azimuthal asymmetry in the
unpolarized Drell-Yan process indicates a substantial non-zero
value for leading twist $T$-odd distribution function
$h_1^\p(x,\kp^2)$ (which is referred to as Boer-Mulders function)
from phenomenological aspects. Theoretically it has been
demonstrated that $h_1^\p$ of the nucleon and the meson can arise
from final or initial state interaction. In this connection, we
have performed a calculation of $h_{1\pi}^\p$ of the pion in a
simple antiquark spectator model by taking into account final
state interaction, and estimated the consequent $\cos2\phi$
azimuthal asymmetry in the unpolarized $\pi^-N$ Drell-Yan process
which is then compared with experimental data measured by NA10
Collaboration. In the calculation we adopt the
pion-quark-antiquark effective coupling as a dipole form factor.
We find that the resulting $h_{1\pi}^\p$, together with $h_{1}^\p$
of the nucleon resulting from a similar treatment with
nucleon-quark-diquark coupling as a dipole form factor, can give a
good agreement of the estimated $\cos2\phi$ azimuthal asymmetry
with experimental data from NA10 Collaboration. This provides a
new indication on the role of $T$-odd distribution $h_{1}^\p$ to
the $\cos2\phi$ asymmetry in the unpolarized Drell-Yan process
from initial state interaction.

{\bf Acknowledgements} We acknowledge the helpful discussion with
Vincenzo Barone. This work is partially supported by National
Natural Science Foundation of China (Nos.10025523, 90103007, and
10421003), by the Key Grant Project of Chinese Ministry of
Education (No.~305001), and by the Italian Ministry of Education,
University and Research (MIUR).


\begin{thebibliography}{99}
\bibitem{bhs02a} S.J. Brodsky, D.S. Hwang, I. Schmidt, Phys. Lett. B 530 (2002) 99.

\bibitem{bhs02b} S.J. Brodsky, D.S. Hwang, I. Schmidt, Nucl. Phys. B 642 (2002) 344.
\bibitem{sivers90} D. Sivers, Phys. Rev. D 41 (1990) 83;\\
D. Sivers, Phys. Rev. D 43 (1991) 261.
\bibitem{abm95} M. Anselmino, M. Boglione, F. Murgia, Phys. Lett. B 362 (1995) 164.
\bibitem{collins93} J.C. Collins, Nucl. Phys. B 396 (1993) 161.
\bibitem{collins02} J.C. Collins,  Phys. Lett.  B 536 (2002) 43.
\bibitem{bjy03} A.V. Belitsky, X. Ji, F. Yuan, Nucl.
Phys. B 656 (2003) 165.
\bibitem{bmp03} D. Boer, P.J. Mulders, F. Pijlman, Nucl. Phys. B 667 (2003) 201.
\bibitem{gg03} L.P. Gamberg, G.R. Goldstein, K.A. Oganessyan, Phys. Rev.
D 67 (2003) 071504(R).
\bibitem{bsy04} A. Bacchetta, A. Sch\"{a}fer,
J.-J. Yang, Phys. Lett. B 578 (2004) 109.
\bibitem{lm04a} Z.~Lu, B.-Q.~Ma, Nucl. Phys. A 741
(2004) 200.
\bibitem{hermes04} HERMES Collaboration, A. Airapetian, et
al., Phys. Rev. Lett. 94 (2005) 012002. 
\bibitem{bm98} D. Boer,
P.J. Mulders, Phys. Rev. D 57 (1998) 5780.
\bibitem{boer99} D. Boer, Phys.
Rev. D 60 (1999) 014012.
\bibitem{na10} NA10 Collaboration, S. Falciano, et al., Z. Phys. C 31 (1986)
513;\\
NA10 Collaboration, M. Guanziroli, et al., Z. Phys. C 37 (1988)
545.
\bibitem{conway89} J.S. Conway, et al., Phys. Rev. D 39 (1989) 92.
\bibitem{bbh03} D. Boer, S.J. Brodsky, D.S. Hwang, Phys.
Rev. D 67 (2003) 054003.
\bibitem{lm04b} Z. Lu, B.-Q. Ma, Phys. Rev. D 70 (2004)
094044.
\bibitem{tm96} P.J. Mulders, R.D.
Tangerman, Nucl. Phys.  B 461 (1996) 197.

\bibitem{jmr97} R. Jakob, P.J. Mulders, J. Rodrigues, Nucl. Phys.
A 626 (1997) 937.
\bibitem{jy02} X. Ji, F. Yuan, Phys. Lett. B 543
(2002) 66.
\bibitem{cs82} J.C. Collins, D.E. Soper, Nucl. Phys. B 194 (1982)
445.
\bibitem{cs77} J.C. Collins, D.E. Soper,
Phys. Rev. D 16 (1977) 2219.
\bibitem{bnm93} A.
Brandenburg, O. Nachtmann, E. Mirkes, Z. Phys. C 60 (1993) 697.
\bibitem{bbkd94} A. Brandenburg, S.J. Brodsky, V.V. Khoze, D.
M\"{u}ller, Phys. Rev. Lett. 73 (1994) 939.
\bibitem{ehvv94} K.J. Eskola, P. Hoyer, M. V\"{a}nttinen,
 R. Vogt, Phys. Lett. B 333 (1994) 526.

\end{thebibliography}
\end{document}